\documentclass[useAMS,usenatbib,usegraphicx]{mn2e}
\usepackage{times}

\newcommand{\ha}{\hbox{H$\alpha$}}
\newcommand{\hb}{\hbox{H$\beta$}}
\newcommand{\hd}{\hbox{H$\delta$}}
\newcommand{\hg}{\hbox{H$\gamma$}}
\newcommand{\oiii}{\hbox{[O\,{\sc iii}]}}
\newcommand{\kms}{\hbox{km~s$^{-1}$}}
\newcommand{\zsp}{\hbox{$z_{\rm sp}$}}
\newcommand{\zph}{\hbox{$z_{\rm ph}$}}
\newcommand{\rth}{\hbox{$r_{\rm 200}$}}

\title{Photometric and Spectroscopic Study of Abell 0671}
\author[Z. Z. Pan et al.]
{Zhizheng Pan$^{1,2}$, Qirong Yuan$^{3}$, Xu
Kong$^{1,2}$\thanks{E-mail:xkong@ustc.edu.cn}, Dongxin Fan$^{4}$,Xu
Zhou$^{5}$,
Xuanbin Lin$^{1,2}$\\
$^{1}$ Center of Astrophysics, University of Science and
Technology of China, Jinzhai Road 96, Hefei 230026, China\\
$^{2}$ Key Laboratory for Research in Galaxies and Cosmology,
USTC, CAS, China\\
 $^{3}$ Department of Physics, Nanjing Normal
University,
WenYuan Road 1, Nanjing 210046, China\\
$^{4}$ Department of physics and Electronics, Guangxi Teachers
Education University, Nanning 530001, China\\
 $^{5}$ National
Astronomical Observatories, Chinese Academy of Sciences, Datun Road
20A, Beijing 100012, China }
\begin{document}
\date{Accepted 0000 December 00. Received 0000 December 00;
in original form 0000 October 00}

%\pagerange{\pageref{firstpage}--\pageref{lastpage}} \pubyear{2011}
\maketitle \label{firstpage}

\begin{abstract}
In this paper we present a photometric and spectroscopic study of the
nearby galaxy cluster Abell 0671 (A671) with 15 intermediate-band
filters in the Beijing-Arizona-Taiwan-Connecticut (BATC) system and
the Sloan Digital Sky Survey (SDSS) data. After a
cross-identification between the photometric data obtained from the
BATC and SDSS, a list of 985 galaxies down to $V \sim20.0$ mag in a
view field of 58$\arcmin$$\times$58$\arcmin$ is achieved, including
103 spectroscopically confirmed member galaxies. The photometric
redshift technique is applied to the galaxy sample for further
membership determination. After the color-magnitude relation is taken
into account, 97 galaxies brighter than $h_{\rm BATC}=19.5$ mag are
selected as new member galaxies. Based on the enlarged sample of
cluster galaxies, spatial distribution, dynamics of A671 are
investigated. The substructures of A671 are well shown by the sample
of bright members, but it appears less significant based on the
enlarged sample, which is mainly due to larger uncertainties in the
light-of-sight velocities of the newly-selected faint members. The
SDSS r-band luminosity function of A671 is flat at faint magnitudes,
with the faint end slope parameter $\alpha$=-1.12. The SDSS spectra
allow us to investigate the star formation history of bright cluster
galaxies, and the galaxies in the core region are found to be older
than those in the outskirts. No environmental effect is found for
metallicities of the early-type galaxies (ETGs). Both mean stellar
ages and metallicities in bright member galaxies are found to be
correlated strongly with their stellar masses assembled, and such
correlations are dependent upon morphology. The possitive correlation
between age and stellar mass supports the downsizing scenario. By
comparing ETG absorption-line indices with the state-of-art stellar
population models, we derive the relevant parameters of simple
stellar population (such as age, [Fe/H], [Mg/Fe], [C/Fe], [N/Fe], and
[Ca/Fe]). The ETGs at cluster center tend to have smaller \hb\
indices, indicating that central ETGs are likely to be older. The
distribution of total metallicity indicator, [MgFe]$'$, does not show
any environmental effects. The relations between the simple stellar
population parameters and velocity dispersion in A671 are in good
agreement with previous studies.

\end{abstract}
\begin{keywords}
galaxies: clusters: individual (A671) -- galaxies: distance and
redshifts -- galaxies: kinematics and dynamics -- galaxies: evolution
-- methods: data analysis
\end{keywords}

\section{Introduction}
According to the hierarchical scenario of structure formation,
massive clusters form by merging small groups continuously and
accreting field galaxies along the filament \citep{West 1991,West
1995, Colberg 2000}. Optical cluster surveys reveal that many galaxy
clusters have evidence for dynamically bound substructures
\citep{Rhee 1991,Beers 1991}. A significant fraction
($\sim$40\%-50\%) of clusters show multiple peaks or irregular
surface brightness distribution in the X-ray images, indicating that
they are still at dynamically active stage, far from equilibrium
\citep{Jones 1999,Schuecker 2001}. Compared with the Einstein-de
Sitter case, clusters in early-epoch universe are expected to be more
relaxed and less substructured, as supported by many N-body
simulation works\citep{Crone 1996,Thomas 1998}. The fraction of
substructured clusters at different redshifts is thus a useful
statistical quantity directly relevant to cosmology. Studies on the
dynamics of galaxy clusters thus provide a unique tool to put
constraint on the models of cluster formation and evolution.

Dense environment in galaxy clusters should have produced influence
on physical properties and evolutionary path for the member galaxies.
Previous studies have found that the observational properties of
galaxies correlate strongly with local galaxy environment
\citep{Gomez 2003,Kauffmann 2004,Baldry 2006}. One of the most
well-studied relation in galaxy clusters is the morphology-density
relation \citep{Dressler 1980,Postman 1984,Whitmore 1991,Goto
2003,Holden 2007}. The core region of a cluster is usually dominated
by early-type galaxies (ETGs), while the outer region is dominated by
late-type galaxies (LTGs). It is well appreciated that the LTGs
gradually lost their gas reservoirs when they were accreted into the
core region, and finally evolved into lenticular galaxies (S0). This
picture of morphology evolution in galaxy clusters has been supported
in high-z morphology-density relation studies \citep{Dressler
1997,Fasano 2000,Smith 2005,Postman 2005}. However, it is still
uncertain and controversial how local galaxy environment affects star
formation histories of cluster galaxies.
% \citep{Poggianti 2004}.

The Beijing-Arizona-Taiwan-Connecticut (BATC) system has spent much
time in observing a sample of more than 30 nearby ($z<0.1$) galaxy
clusters at different dynamic statuses, aiming at studying their
dynamic substructures, luminosity functions, and the star formation
properties of cluster galaxies. Abell 0671 (A671; z=0.0502) is one
target of the BATC galaxy cluster survey. Its Abell richness $R$ is
set to be 0 \citep{abell58}, with Bautz-Mogan type II-III
\citep{bm70}. The X-ray emission from the cluster center has been
detected by the \emph{Einstein} observatory and the \emph{ROSAT}
All-Sky Survey (RASS). The X-ray luminosity of A671 detected in the
RASS 0.1-2.4 keV band is $0.9\times10^{44}\textmd{erg}$
$\textmd{s}^{-1}$, and the X-ray temperature is 3.1 keV
\citep{Ebeling 1998}, which confirms that this cluster is a
relatively poor system. Figure 1 shows the smoothed contours of the
{\em Einstein} X-ray image and the radio map at 1.4 GHz from the NRAO
VLA Sky Survey (NVSS), superimposed on the optical image in the
BATC$-h$ band. No radio emission is detected at the center of A671.
The X-ray surface density contour is quite regular with a single
symmetric peak as reported by \citet{Jones 1999}. However, the detail
structure of A671 is possibly blurred due to the low resolution
($\sim$ 1 arcmin spatial resolution) and the large PSF (FWHM
$\sim$1.5 armin) of Einstein IPC image . It is easily seen that the
X-ray emission peak does not coincide with the central brightest
galaxy, IC~2378, with a positional offset of about 90 kpc. A671 is
included in the cluster sample of the WIde-field Nearby
Galaxy-cluster Survey (WINGS), and two substructures in A671 have
been found by \citet{Ramella 2007} recently. For a better
understanding of the dynamics of A671, it is important to construct a
large sample of member galaxies, and the faint galaxies ($18.0<
m_{\rm v}<19.5$) should be taken into account. In this paper, we
present a multicolor photometry of the galaxies in A671 region with
the Beijing-Arizona-Taiwan-Connecticut system. We try to enlarge the
sample of cluster galaxies by applying the photometric redshift
technique to the spectral energy distributions (SEDs) of the
BATC-detected faint galaxies. Based on the SDSS spectra of bright
member galaxies, we will derive the star formation histories and
chemical abundances, and try to find any clues of environmental
effects on the physical parameters of cluster galaxies.

\begin{figure}
\centering
\includegraphics[width=85mm,angle=0]{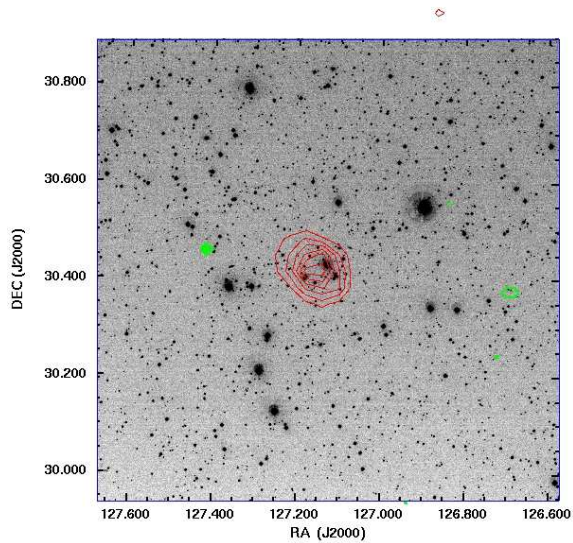}
\caption{
The smoothed contours of $Einstein$ image ($0.5-4.5$ keV band)
(red line) and the NVSS map at 1.4 GHz (green line), superimposed
on the BATC$-h$ band image.
The sizes of gaussian smoothing windows are adopted as $30\arcsec$
and $1.2\arcmin$ for radio and X-ray contours, respectively.}
\end{figure}

This paper is organized as follows:  we present the BATC photometric
observations and data reduction in Section\,2. In Section\,3, we
analyze the galaxies with known spectroscopic redshifts in the A671
field. In Section\,4, we apply photometric redshift technique to
select faint member galaxies in A671. In Section\,5, dynamic
substructures and luminosity function are investigated based on the
enlarged sample of member galaxies. In Section\,6, we derive the star
formation histories and chemical abundances of the ETGs in A671.
Finally, we summarize our work in Section\,7. Throughout this paper,
we assume the cosmological parameters as $\Omega_{\rm m}=0.3$,
$\Omega_{\rm \Lambda}=0.7$, $H_{\rm 0}=70$ \kms Mpc$^{-1}$.

\section{Observation and data reduction}

The BATC survey is based on the 60/90 $cm$ $f/3$ Schmidt Telescope of
National Astronomical Observatories, Chinese Academy of Science
(NAOC), located at Xinglong Station. The BATC system contains 15
intermediate-band filters, covering a wavelength from 3000 to 10000
\AA, which are designed to avoid night sky emission lines \citep{Fan
1996,Kong 2000}. The transmission curves of BATC filters can be found
in \citet{Fan 1996}. Before October 2006, a Ford CCD camera with a
format of $2048\times2048$ was mounted on the telescope, and
photometric observations in 12 bands, from $d$ to $p$, were carried
out. The viewing field was about $58\arcmin\times58\arcmin$, with a
scale of $1\arcsec.7/$pixel. For pursuing better spatial resolution
and higher sensitivity in three blue bands, $a-c$, a new E2V CCD with
$4096\times4096$ pixels was then equipped. The field of view becomes
larger ($92\arcmin\times92\arcmin$) with a spatial scale of
$1\arcsec.35/$pixel. The newly equipped CCD camera has a high quantum
efficiency of 92.2\%.

From March 2003 to October 2007, we totally accumulated 50 hours
exposure for A671 with 15 filters (see the observational information
in Table 1). With an automatic data-processing software, PIPELINE I
\citep{Fan 1996}, we carry out the standard procedures of bias
subtraction, flat-field correction, and position calibration. The
technique of integral pixel shifting has been used in the image
combination during which cosmic rays and bad pixels are removed by
comparing multiple images.

\begin{table*}
 \centering
 \begin{minipage}{140mm}
  \caption{The detail of the BATC filters and observation information of A671}
 \begin{tabular}{@{}ccccccccc}
 \hline
  Number &Filter &$\lambda_{eff} $&FWHM &Exposure &Number of
  &Seeing$^a$
  &Objects &Limiting\\
 &name&$({\textbf{\AA}})$&$({\textbf{\AA}})$&(second)
 &Images&(arcsec)&Detected&mag\\
\hline
1   &a  &3369   &222    &18000  &15 &3.79   &3945   &21.5\\
2   &b  &3921   &291    &7200   &6  &4.90   &4158   &21.0\\
3   &c  &4205   &309    &12600  &12 &4.69   &5738   &21.0\\
4   &d  &4550   &332    &19500  &17 &4.61   &4739   &20.5\\
5   &e  &4920   &374    &15000  &14 &3.78   &5453   &20.5\\
6   &f  &5270   &344    &13800  &13 &4.40   &5466   &20.0\\
7   &g  &5795   &289    &7800   &8  &4.12   &5506   &20.0\\
8   &h  &6075   &308    &7500   &7  &5.62   &5723   &20.0\\
9   &i  &6660   &491    &5460   &7  &3.51   &6675   &20.0\\
10  &j  &7050   &238    &7500   &7  &4.14   &6017   &19.5\\
11  &k  &7490   &192    &12600  &12 &4.09   &5968   &19.5\\
12  &m  &8020   &255    &11100  &10 &4.74   &6176   &19.0\\
13  &n  &8480   &167    &11100  &10 &4.24   &5893   &19.0\\
14  &o  &9190   &247    &15000  &14 &4.13   &5846   &18.5\\
15  &p  &9745   &275    &18600  &16 &4.46   &5046   &18.5\\
 \hline
\end{tabular}
\parbox{80mm} {$^a$ seeing of the combined image}
\end{minipage}
\end{table*}

For detecting and measuring the flux of the sources within a given
aperture in the BATC images, we convert the $a-c$ combined images to
make the pixel size identical with $d-p$ images. We adopt a radius of
4 pixels as a photometric aperture for all the BATC images to the
sources detected by the \emph{SExtractor} codes \citep{Bertin 1996}.
The flux calibration of SEDs is performed by using the Oke-Gunn
\citep{Gunn 1983} standard stars (HD19445, HD84937, BD+26d2606 and
BD+17d4708) which were observed during photometric nights. The
detailed information about calibration can be found in \citet{Zhou
2001}. Because we have no calibration images for $a$, $b$, $c$
filters, we instead perform the model calibration that has been
developed specially for the large-field photometric system by
\citet{Zhou 1999}. As a result, the SEDs of 6782 sources have been
obtained in our catalog. By cross-identifying the BATC sources with
the SDSS photometric data within a search circle (defined as a circle
of 1.5 arcsec), all sources are classified into galaxies and
stars. As a result, 985 galaxies brighter than the BATC$-h$ band
magnitude limit are found by both surveys, which offers a sample for
further analysis.

\section{Analysis of galaxies with known spectroscopic redshifts}

\subsection{Distribution of spectroscopic redshifts}

For studying the dynamics of galaxy cluster A671, 205 galaxies with
known redshifts in our viewing field are extracted from the SDSS DR8
galaxy catalog. Figure 2 shows the distribution of spectroscopic
redshifts of these galaxies. The main concentration with a peak at
$z\sim0.05$ is isolated and less contaminated. There are 103 galaxies
with $0.04<z<0.06$ and they are selected as the member galaxies of
A671, to which we refer as Sample I. To characterize the velocity
distribution, we convert the spectroscopic redshifts (\zsp) into
the rest-frame velocities ($v$) by
$v=c\times(\zsp-\bar{z}_{\rm c})/(1+\bar{z}_{\rm c})$,
where $c$ is the light speed, $\bar{z}_{\rm c}$ is the cluster
redshift with respect to the cosmic background radiation. We take
the NED-given cluster redshift $\bar{z}_{\rm c}$ =0.0502 for A671.
The velocity distribution can be fitted by a Gaussian with a
dispersion of $\sigma=625$ \kms, and it is significantly deviated
from a standard Gaussian function.

To qualify the distribution of radial velocities of member galaxies,
we use the ROSTAT software \citep{Beers 1990} to calculate two
resistant and robust estimators, biweight location ($C_{\rm BI}$) and
scale ($S_{\rm BI}$), analogous to the velocity mean and the standard
deviation. For these 103 galaxies, we achieve $C_{\rm
BI}=14561_{-82}^{+100}$ \kms, $S_{BI}=820_{-50}^{+57}$ \kms. The
errors are determined from the 68\% confidence limits based on 10,000
bootstrp resamplings of the velocity data. Taking a cosmological
correction factor of $(1+z)^{-1}$ into account, the velocity
dispersion of A671 should be $780_{-47}^{+54}$ \kms. \citet{Aguerri
2007} studied a sample of 88 nearby clusters. Only 72 galaxies were
included in their sample, and a smaller $S_{\rm BI}$ was derived.
They found A671 having $C_{\rm BI}=14599^{+19}_{-33}$ \kms and
$S_{\rm BI}=610^{+37}_{-33}$ \kms. Our statistics is based on a
larger sample, thus is more reliable. Both studies confirm that A671
has a comparatively small velocity dispersion.

\begin{figure}
\centering
\includegraphics[width=85mm,angle=0]{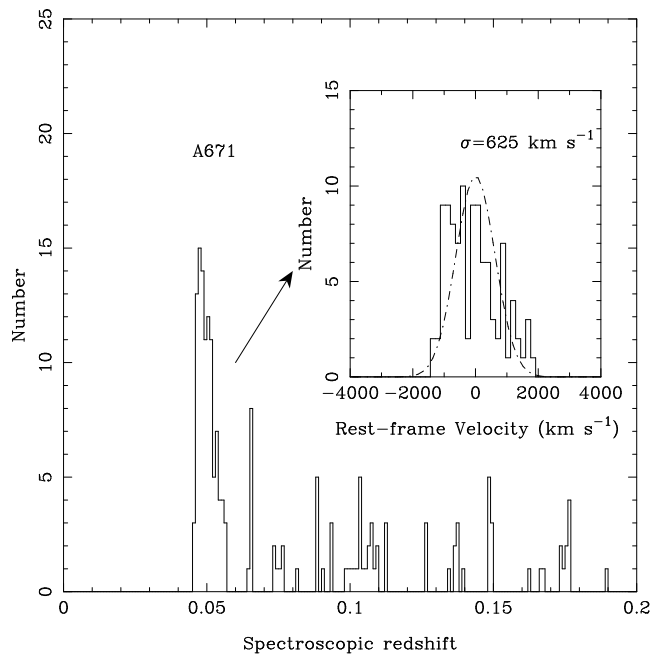}
\caption{
Distribution of the redshifts for 205 galaxies in the A671 field,
with bin size $\Delta z=0.001$. The galaxies in A671 are
centered at $z\sim0.05$. We select the galaxies with $0.04<z<0.06$ as
member galaxies. The smaller panel shows the distribution of
rest-frame velocities of member galaxies in detail, and dash line
represents a gaussian fit to the histogram. }
\end{figure}

\begin{figure*}
\centering
\includegraphics[width=160mm,angle=0]{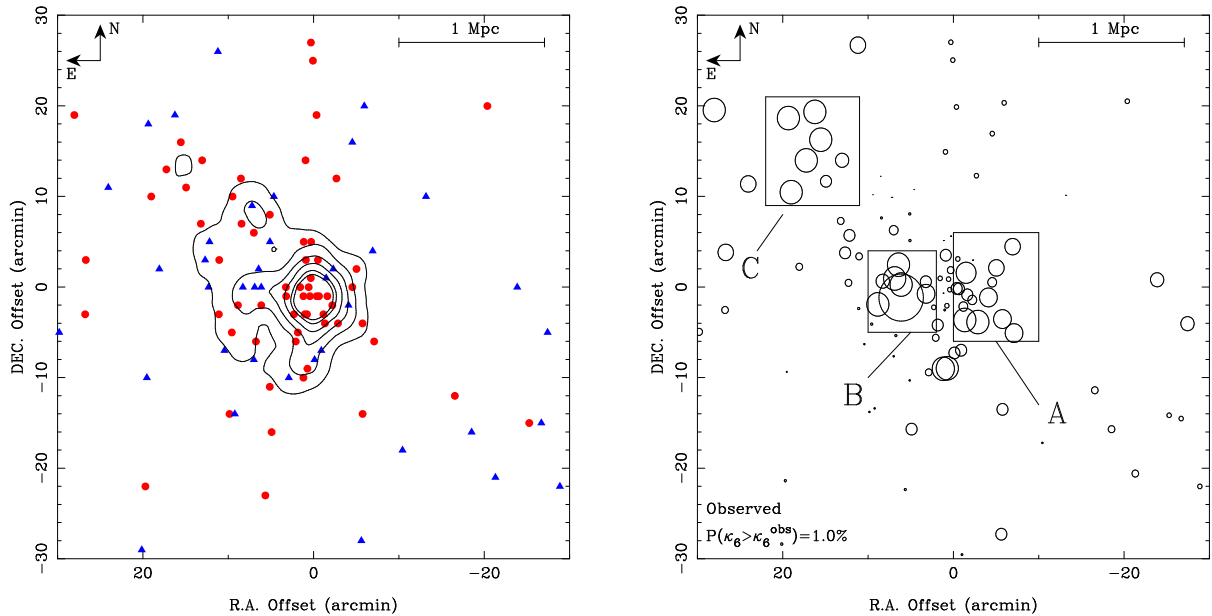}
\caption{ Left: Spatial distribution of 103 known member galaxies of
A671, including 63 early-type galaxies (red circles), and 40
late-type galaxies (blue triangles). The contour map of surface
density for all these galaxies, smoothed by a Gaussian window with
$\sigma=1.6$ arcmin. The contour levels are 0.09,0.15,0.21,0.27,0.33
and 0.39 arcmin$^{-2}$, respectively. Right: Bubble plot for groups
of six nearest neighbors, showing the localized variation in velocity
distribution.}
\end{figure*}

\subsection{Spatial distribution and localized velocity structure}

Because A671 is a nearby cluster, our BATC viewing field can not
cover whole cluster region.
Our photometry focus on a central field of $3.4\times3.4$ Mpc$^2$.
The radius of galaxy cluster, \rth, is defined in former studies
as the boundary of a cluster, within which the mean inner density
is $200\rho_{\rm c}$, where $\rho_{\rm c}$ is the critical density
of the Universe \citep{Gott 1972}.
We calculate the \rth\ for A671 following the formula suggested
by \citet{Carlberg 1997}. The \rth\ is a function of velocity
dispersion. By applying the $S_{\rm BI}$ we derived, the \rth\ of
A671 is 2.25 Mpc, corresponding a slightly larger area than our
viewing field.

\begin{figure}
\centering
\includegraphics[width=80mm,angle=0]{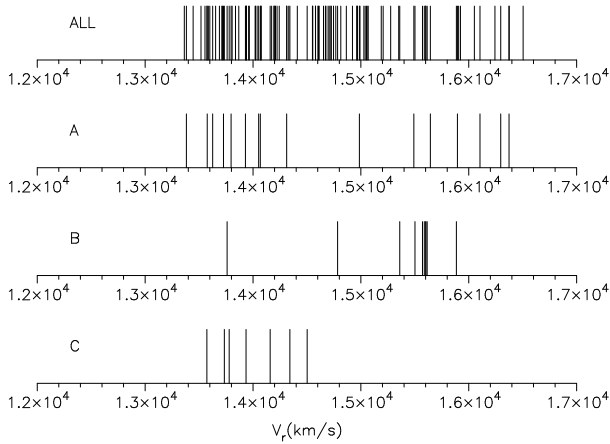}
\caption{ Stripe density plot of velocities of the spectroscopically
confirmed galaxies in whole cluster, clumps A, B, and C,
respectively.}
\end{figure}
Before studying the dynamic structure of A671, we try to classify the
103 known members into ETGs and LTGs. The early-type members should
meet two requirements: (1)with no evident emission lines;(2)with no
evident galaxy arms. We firstly extract those galaxies with
EW(\ha)$<5$ \AA\ as early-type candidates. The EW(\ha) values are
taken from the MPA/JHU
catalog\footnote{http://www.mpa-garching.mpg.de/SDSS/DR7} of SDSS
galaxies. Then we inspect their SDSS-given images and removing those
with arms. As a result, the 103 galaxies are classified into 63 ETGs
and 40 LTGs.

The left panel of Figure 3 presents the spatial distribution of 103
known member galaxies within our field of view, with the central
position of A671, R.A.=$8^{h}28^{m}29^{s}$,
Dec.=$30{\degr}25{\arcmin}01{\arcsec}$ (for the J2000 equinox). We
superpose the contour map of surface density that has been smoothed
by a Gaussian window with ${\sigma}=1.6{\arcmin}$. As shown in Figure
3, the member galaxies are mainly concentrated in the central region
within a radius of 1 Mpc. The irregular contour in the east and north
corresponds to the two substructures found by \citet{Ramella 2007}.
More statistical tests should be performed before we can reach a firm
conclusion that there are substructures in A671.

To show the substructures of A671 in both velocity space and
projected map, we make use of the $\kappa$-test \citep{Colless 1996}
for the 103 galaxies. The statistic variable, $\kappa_{\rm n}$, is
defined to quantify the local deviation on the scale of groups of $n$
nearest neighbors. A larger $\kappa_{\rm n}$ indicates a greater
probability that the local velocity distribution differs from overall
velocity distribution.
The probability $P(\kappa_{\rm n}>\kappa_{\rm n}^{\rm obs})$ can be
calculated by Monte Carlo simulations with random shuffling velocities.
When the scale of the nearest neighbors $n$ varies from 3 to 9, the
probabilities $P(\kappa_{\rm n}>\kappa_{\rm n}^{\rm obs})$ are nearly
zero, which means the substructure appears very obvious at different
scales.
The bubble plot at the scale of $n=6$ is given in the right panel of
Figure 3.
Since the bubble size is proportional to
$-{\rm log} P(D_{\rm n}>D_{\rm n}^{\rm obs})$, the clustering of
large bubbles is a good tracer of dynamical substructure.

As we can see in the bubble plot, the central region of A671 is
dominated by two clumps of bubbles, and a clump of large bubbles in
the north-east is also remarkable. We refer to these three clumps as
A, B, and C, which contain 16, 10, 8 galaxies, respectively. For
confirming whether these clumps trace the real substructures, we
present the velocity distributions for subsamples A, B, and C in
Figure 4, as well as the velocity distribution of the whole sample.
It is easily seen that the velocity distributions for the three
clumps are indeed deviated from that of the whole sample. The mean
velocities of the subsamples A, B, and C are  $14638_{-641}^{+408}$,
$15567_{-266}^{+23}$, and $13992_{-174}^{+171}$ \kms, respectively.
The mean velocities of subsamples B and C are significantly deviated
from the mean velocity of the whole sample, 14561 \kms. Though the
mean velocity in clump A is similar to that of the whole sample, the
galaxies in clump A have a remarkable bimodal velocity distribution,
which may imply the existence of two different groups. To quantify
the bimodality significance, we apply the Gaussian mixture modeling
(GMM) method \citep{Muratov 2010} to the velocity distribution for
clump A. The result shows that clump A consists of two components,
which peaked at 13829 \kms and 15824 \kms. The unimodal velocity
distribution can be rejected at 99\% significance level.

Although the velocity distribution of the subsamples are
significantly deviated from the whole sample, their masses should
also be large enough if they are real substructures. The masses of
the A671 and its clumps can be estimated by applying the virial
theorem. Assuming that each subcluster is bound and the galaxy orbits
are random, the virial mass ($M_{v}$) can be derived from the
following standard formula \citep{Geller 1973,Oegerle 1994}:
\begin{equation}
M_v=\frac{3\pi}{G}\sigma_{r}^{2}DN_{p}(\sum^{N}_{i>j}\frac{1}{\theta_{ij}})^{-1},
\end{equation}
where $\sigma_{r}$ is the light-of-sight velocity dispersion, $D$ is
the cosmological distance of the cluster, $N_{p}=N(N-1)/2$ is the
number of galaxy pairs, and $\theta_{ij}$ is the angular distance
between galaxies $i$ and $j$. The viral masses of A671 is
$9.49\times10^{14}M_{\odot}$. Clump B is the most remarkable among
the 3 subsamples, with viral mass of $1.53\times10^{14}M_{\odot}$.
The A clump consists of two groups, with viral mass of $\sim$
$7.0\times10^{13}M_{\odot}$ for each. Clump C has similar viral mass.
Thus we conclude that A671 is not a simple relaxed cluster, but most
likely at a dynamically active stage.

\section{SED Selection Of Faint Cluster Galaxies}

The technique of photometric redshift can be used to estimate the
redshifts of galaxies by using the SED information covering a wide
range of wavelength instead of spectroscopy. This technique has been
extensively applied to the multicolor photometric surveys for
detecting the faint and distant galaxies, and for subsequential
selection of cluster galaxies \citep{Fernadez 1999,Ilbert 2009, Kong
2009}. Based on the standard SED-fitting code called $HyperZ$
\citep{Bolzonella 2000}, for a given object, the photometric
redshift, \zph, corresponds to the best fit (in the $\chi^{2}$
-sense) of its photometric SED with the template SED generated by
convolving the galaxy spectra in the template library with the
transmission curves of specified filters. Previous work has
demonstrated the accuracy of photometric redshift with the BATC
multicolor photometric data \citep{Xia 2002,Yang 2004,Liu 2011,Zhang
2011}. In our SED fitting, only normal galaxies are taken into
account in the reference templates. Dust extinction with a reddening
law of the Milky Way \citep{Allen 1976} is adopted, and $A_{\rm V}$
is allowed to be flexible in a range from 0.0 to 0.5, with a step of
0.05. The photometric redshift of a given galaxy is searched from 0.0
to 0.6, with a step of 0.005. We apply this technique to all BATC
galaxies brighter than $h_{\rm BATC}=19^m.5$. The procedure of
SED-fitting has provided the best-fit photometric redshift and its
uncertainty for each galaxy.

\begin{figure*}
\centering
\includegraphics[width=160mm,angle=0]{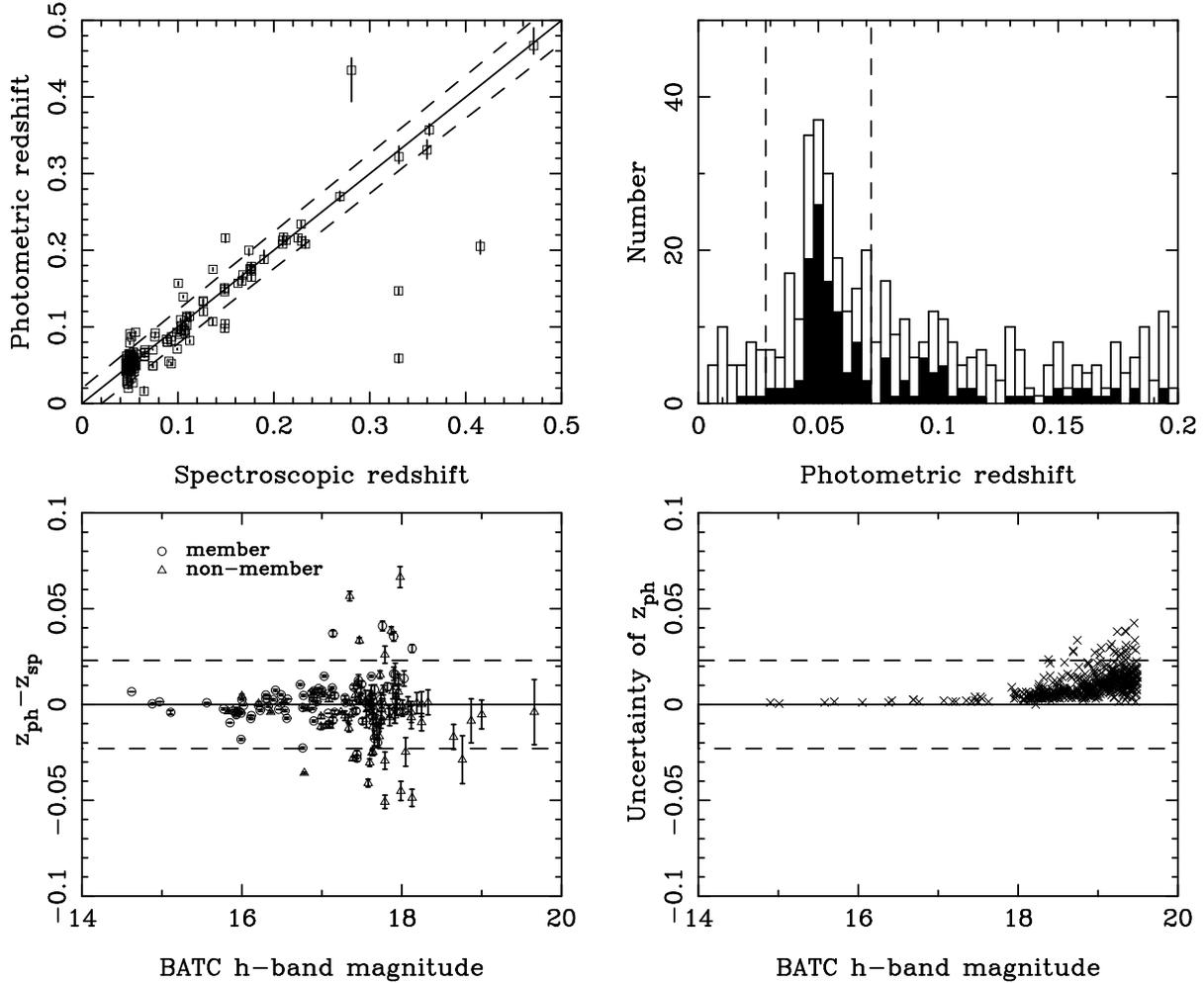}
\caption{Top-left: comparison between \zph\ and \zsp\ 167 galaxies
detected with at least 12 BATC bands. The error bars shows the 68\%
significance coefficient level. Solid line denotes $\zph=\zsp$, and
the dash lines denote the deviation of $0.02(1+z)$. There is no
systematic deviation between the photometric redshifts and the
spectroscopic redshifts. Top-right: \zph\ distribution for the
galaxies down to $h_{\rm BATC}=19^m.5$, with a bin size of $\Delta
z=0.004$. The black histogram shows the \zph\ of the 167 galaxies
with \zsp. The dash line shows our faint member candidates selection
criterion. Bottom-left: difference of \zph\ and \zsp\ for the 167
galaxies with spectroscopic redshifts. Circles denote
member-galaxies, and triangles denote non-members. The dash lines
denote the member selection cataria($2\sigma=0.023$). The error bars
are given by the $HyperZ$ codes, with 68\% confidence level.
Bottom-right: crosses denote $1\sigma$ errors of the of the \zph\ for
the left galaxies with h-band magnitude brighter than 19.5 mag in the
viewfield of A671. }
\end{figure*}

From the 205 galaxies with known spectroscopic redshifts (\zsp),
we select 167 galaxies (including 91 member galaxies) that are
simultaneously detected in at least 12 BATC bands to derive their
\zph\ values. A comparison between \zph\ and \zsp\ values is
shown in the top-left panel of Figure 5. The error bar of \zph\
corresponds to 68\% confidence level in photometric redshift
determination. The solid line denotes $\zph=\zsp$, and the dashed
lines show an average redshift deviation of $0.02(1+z)$. It is
obvious that our \zph\ estimates are basically consistent with
their \zsp\ values. For 91 member galaxies in our spectroscopic
sample, the mean value and standard deviation of their \zph\
values are 0.0505 and 0.0112, respectively. There is no systematic
offset in the \zph\ domain with respective to the \zsp\
distribution. Statistically, 85 member galaxies (about 93 percent)
are found to have their photometric redshifts within $\pm 2\sigma$
deviation, in a range from 0.028 to 0.073, demonstrating the
robustness of our \zph\ estimate. This \zph\ region can be
applied as a selection criterion in the following membership
determination for faint galaxies.

The top-right panel of Figure 5 shows the histogram of photometric
redshifts for all galaxies down to $h_{\rm BATC}=19^m.5$. The black
histogram shows \zph\ distribution for the 167 galaxies with known
spectroscopic redshifts mentioned above. As expected, the peak in the
\zph\ distribution is around z=0.05. In the bottom-left panel of
Figure 5, we show the \zph\ uncertainties for these 167 galaxies as a
function of BATC$-h$ band magnitude. It is remarkable that the \zph\
deviation of fainter galaxies tends to be larger. For the faint
galaxies with $h_{\rm BATC}=18^m.0$, our \zph\ estimate is still
robust, but with larger uncertainty. For the remainder galaxies
without \zsp\ values, we give a plot of their \zph\ uncertainties
versus BATC$-h$ band magnitudes in the bottom-right panel in Figure
5. The larger \zph\ uncertainties for faint galaxies are mainly due
to larger magnitude errors in photometry. For a reliable membership
determination based on the \zph\ estimate, we exclude the galaxies
fainter than $h_{\rm BATC}=19^m.5$, and take the galaxies with
$0.028<\zph<0.073$ as member candidates. Due to the robustness of our
photometric redshift technique, it is conservative that our selecting
criterion would be able to select 80-90\% of faint members with least
contaminants.

\begin{figure*}
\centering
\includegraphics[width=160mm,angle=0]{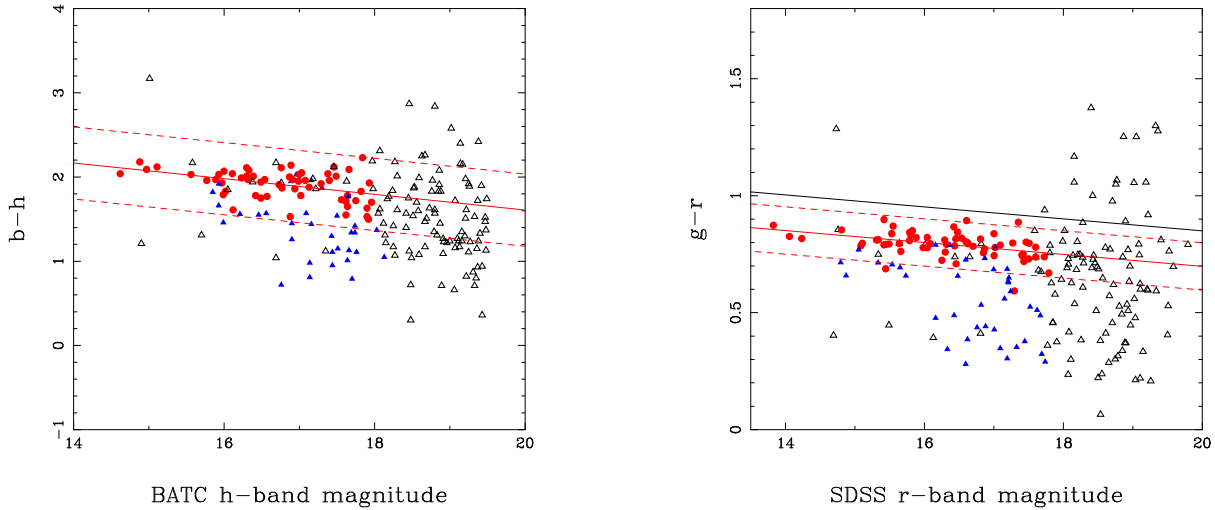}
\caption{
The left panel shows the BATC C-M relation, using color index $b-h$
and $h-$band magnitude, for the spectroscopically confirmed members
and the galaxies with $0.028<\zph<0.073$.
The right panel shows the SDSS C-M relation, using color index $g-r$
 and SDSS $r-$band model magnitude. The filled red circles denote the
spectroscopically confirmed ETGs in A671. The filled triangles denote
the spectroscopically confirmed LTGs in A671. The open triangles
denote the newly selected galaxies with $0.028<\zph<0.073$. The solid
red lines represent the subsequence fitted with the known early-type
members. The dashed red lines represent the $1 \sigma$ of the
best-fit C-M relations. The black solid line represents our selection
criterium based on the SDSS C-M relation.}
\end{figure*}

It is well known that there exits a correlation between color and
absolute magnitude for the ETGs (C-M relation) \citep{Bower 1992}, in
the sense that brighter ETGs appear redder, which can be used for
verifying the membership selection of the ETGs. The left panel of
Figure 6 presents the correlation between the color index, $b-h$, and
BATC$-h$ band magnitude for the whole member candidates down to
$h_{\rm BATC}=19^m.5$, while the right panel shows the SDSS color
index, $g-r$, and $r-$band magnitude.
The diagrams include the following categories of sources :
(1) spectroscopically confirmed early-type member candidates (denoted
by filled circles),
(2) spectroscopically confirmed late-type member candidates (denoted
by filled triangles),
(3) newly-selected faint member candidates (denoted by open
triangles). The solid line represents the linear fitting with the 63
spectroscopically confirmed ETGs:
$b-h=-0.09(\pm0.02)h+3.47(\pm0.43)$, and the dashed line represents
$1 \sigma$ deviation. The linear fit of the ETGs in the right panel of
Figure 6 is $g-r=-0.026(\pm0.006)r+1.21(\pm0.10)$. As shown in both
figures, the early-type member galaxies follow a very tight
color-magnitude relation, faint member candidates also follow the
same C-M relation basically, but seem to be more scattered. This
might be caused by some high-$z$ galaxies that have been mixed into
our faint member candidates. For excluding these contaminants, we
utilize the SDSS C-M relation, and remove the ETG candidates with
color indices $g-r$ 0.15 mag redder than the red sequence denoted by
the black solid line.

Finally, we obtain a list of 97 newly-selected member galaxies.
Combined with those 103 spectroscopically confirmed members, we form
an enlarged sample of 200 galaxies in A671, to which we refer as
Sample II in following investigation.

\section{The Physical Properties of A671}

\subsection{The Spatial Distribution and Velocity Structure}

The left panel of Figure 7 shows the projected positions of the
galaxies in Sample II, superposed with the contour map of surface
density smoothed by a Gaussian window with $\sigma=1.6\arcmin$.
The 103 member galaxies with known spectroscopic redshifts are
denoted by filled symbols, and 97 photometrically selected galaxies
are denoted by open symbols. Red symbols represent the early-type
member galaxies and blue symbols represent late-type ones.
Our BATC multicolor photometry facilitates the finding of large
number of faint member galaxies, and makes the underlying
substructure along north-east direction more remarkable.
Basically, the distribution of faint galaxies traces that of
bright ones, and no significant substructures are found with Sample
II.

\begin{figure*}
\centering
\includegraphics[width=160mm,angle=0]{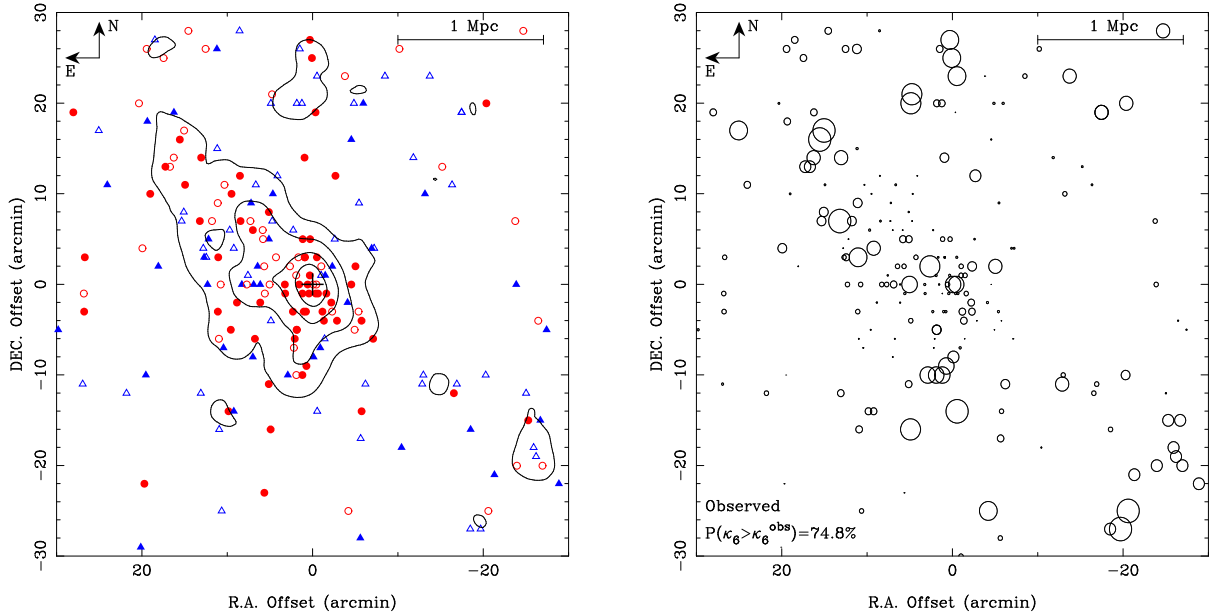}
\caption{
Left: spatial distribution of all member galaxies in Sample
II. Filled symbols denote the galaxies with known spectroscopic
redshifts, while open symbols represent newly selected members.
Red circles denote red-sequence galaxies, while blue open
triangles denote those bel1ow $1 \sigma$ of the red-sequence.
Right: bubble map shows the local velocity distribution for groups
of six nearest neighbors for the whole member galaxies.}
\end{figure*}

The morphology segregation becomes more remarkable in Sample II. Both
bright and faint ETGs are highly concentrated in the core region,
while the LTGs are scattered in the outskirts. The shape of contour
map seems in accord with the X-ray image, which might demonstrate the
reliability of our membership selection based on the BATC multicolor
photometry.

For detecting the potential substructures in A671, we perform the
$\kappa$-test for Sample II. The right panel of Figure 7 shows the
bubble plot that characterizes the degree of difference between the
localized velocity distribution, for groups of six nearest neighbors,
and the overall velocity distribution. We performed $10^3$
simulations to estimate probability
$P(\kappa_{\rm n}>\kappa_{\rm n}^{\rm obs})$ for
different group sizes. The probability is found to be more than 5\%
in all cases, which means that no substructure is detected at
$2 \sigma$ significance (see Table 2). This is not consistent with the
conclusion that we have achieved based on Sample I. We think that the
substructure unveiled by the spectroscopic redshifts is surely true.
Above inconsistency can be well interpreted that the velocities
derived from the \zph\ estimates are not accurate enough to
reflect the subtle velocity structure. The abnormity in velocity
distribution of substructures might have been smoothed/swept by the
\zph\ uncertainties of 97 newly selected faint galaxies. So the
$\kappa$-test on Sample II might be misleading.
Follow-up spectroscopy of these faint member galaxies are needed
if one wants to investigate the substructures in A671 in detail.

\begin{table}
\centering
 \caption{Result of the $\kappa$-test for member galaxies}
 \begin{tabular}{ccc}
 \hline \noalign{\smallskip}
 Neighbors & $P(\kappa_{\rm n}>\kappa_{\rm n}^{\rm obs})$ & $P(\kappa_{\rm n}>\kappa_{\rm n}^{\rm obs})$\\
 size ($n$) & Sample I & Sample II \\
\hline\noalign{\smallskip}
 3 &1.4\% &34.7\% \\
 4 &0.6\% &35.9\% \\
 5 &2.8\% &65.3\% \\
 6 &1.0\% &74.8\% \\
 7 &0.5\% &71.5\% \\
 8 &0.7\% &65.7\% \\
 9 &1.4\% &62.0\%\\
   \noalign{\smallskip}
 \hline
\end{tabular}
\end{table}

\subsection{The Luminosity Function}

The luminosity function (LF) is a key diagnostic for clusters because
it is tightly related to dynamical evolution and merging history of
galaxy clusters. The LF has been widely studied in the past decades
and it is well described by the Schechter function \citep{Schechter
1976}:
\begin{equation}
\phi(L)dL=\phi^*\,(L/L^*)^{\alpha}exp(-L/L^*)\,d(L/L^*),
\end{equation}
where $\phi^*$, $L^*$, and $\alpha$ are the normalization parameter,
 the characteristic luminosity, and the slope parameter at faint end,
respectively. In the domain of absolute magnitude, the Schechter
function can be expressed as:
\begin{equation}
\phi(M)dM=\phi^*10^{0.4(\alpha+1)(M^*-M)}exp[-10^{0.4(M^*-M)}]dM,
\end{equation}
where $M^*$ is the characteristic absolute magnitude.

\begin{figure}
\centering
\includegraphics[width=80mm,angle=0]{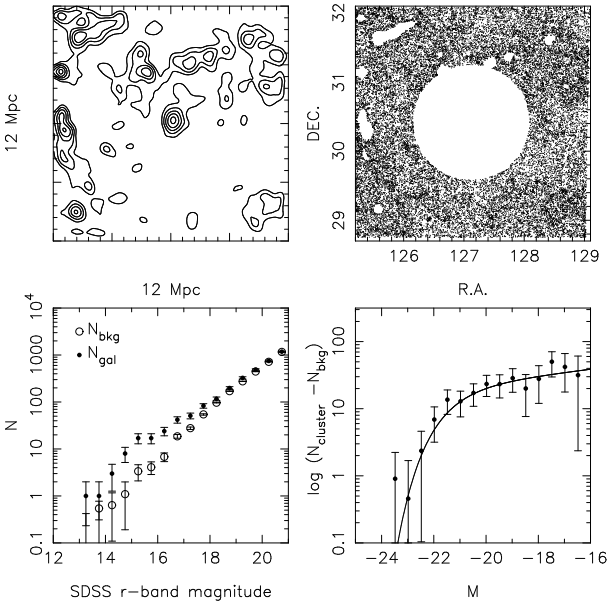}
\caption{Top left: Density map of a $12\times12$ Mpc region around
A671. Contour levels are from 1$\sigma$ to 5$\sigma$, spaced
1$\sigma$ apart. Top right: The region which is used to estimate the
local background. Dots denote the galaxies with the SDSS r-band
magnitude brighter than 21.0 mag. Empty regions represent the removed
overdensity area. Bottom left: galaxy number counts extracted within
r=1.7 Mpc region (filled circles), and those in the local background
(open circles) for A671. Bottom right: background-subtracted number
counts for A671. The solid line represents the best-fit Schechter
function. }
\end{figure}

The most challenging issue in measuring the LFs of galaxy clusters
(GCs) is that one needs to pick up cluster members out of background
galaxies along the line of sight. Ideally, one needs the
spectroscopic redshifts for all galaxies to exclude non-members in
the cluster field. Unfortunately spectroscopic measurements are
rather time-consuming. Our BATC photometry enables us to select a set
of cluster member candidates by utilizing the photometric redshift
technique. As seen in Figure 5, the accuracy of \zph\ is a function
of galaxy apparent magnitude. Further corrections are still needed to
remove the contribution of contaminant sources and compensate the
missing members when investigating the LFs. Unfortunately the exact
form of the correction function, particularly at the faint
magnitudes, is very difficult to be derived.

Using the SDSS r-band photometric data, we perform the statistical
background subtraction to estimate the contribution of non-members to
the number counts of galaxies in the cluster direction, by measuring
the projected number counts of field galaxies outside the cluster
region. The background is estimated with a $12\times12$ Mpc region
centered on the cluster centroid , outside the cluster region defined
by a radius of 3 Mpc, where the contamination from cluster galaxies
should be negligible. Following the method of \citet{Paolillo 2001},
we firstly generate a density map of galaxies in the background
region by convolving the projected distribution of galaxies with a
Gaussian kernel of $\sigma$=250 kpc in the cluster rest frame (the
typical size of a cluster core). Then we mask out all density peaks
which are above 3$\sigma$ level from the background region.
Masked-out regions covered about 3.3\% of the whole background area.
Finally we calculate the number counts from the remain galaxies to
estimate the background number counts in the cluster direction.

Recent studies have provided evidences that the LFs of GCs do vary
with clustercentric radius \citep{Beijersbergen 2002,Hansen 2005}. A
suitable region should be chosen which is large enough to contain
most member galaxies and do not include much projected contamination.
We adopt a aperture of r=30\arcmin centered at the cluster centroid
(about 1.7 Mpc at the rest frame of A671). The results of the
background estimation and final luminosity function are showed in
Figure 8. The apparent magnitudes are converted to absolute
magnitudes by the relation
\begin{equation}
M=m-DM(z)-K_{0.1}(z)
\end{equation}
Where $DM(z)$ is the distance modulus as determined from the redshift
assuming a particular cosmology. $K_{0.1}(z)$ is the $K$-correction
from a galaxy at $z$ to $z=0.1$. We estimate the $K_{0.1}(z)$ using
the software \texttt{KCORRECT}(version $4_{-}1_{-}4$, \citet{Blanton
2003}). As showed in the bottom-right panel of Figure 8, a single
Schechter function can fit the data very well. The best fit
parameters are $\phi^*$=21.0, $M^*$=-21.6, $\alpha$=-1.12, which is
in good agreement with \citet{de Filippis 2011}. No strong ``
upturn'' at faint magnitudes is observed for A671.

%Recent studies provided evidences that LFs of GCs does vary with
%clustercentric radius \citep{Beijersbergen 2002,Hansen 2005}.
%We firstly transform the BATC magnitudes of member galaxies into the
%5conventional Kron-Cousins $B$, $V$ and $R$ magnitudes via the
%equations given in \citet{Zhou 2003}. Figure 8 presents the LF of
%galaxies in A671. Due to the potential incompleteness at faint end,
%the data points, denoted by filled symbols, are not taken into
%account during the fitting procedure. The solid line is the best-fit
%Schechter function, and the derived parameters are given in Table 3.

%In some clusters, a single Schechter function can not perfectly
%fitting the data, for example, A963 \citep{Driver 1994}, A2554
%\citep{Smith 1997}, A168 \citep{Yang 2004}. However, as shown in
%Figure 8, a single Schechter function can be used to fit the LFs
%of A671 well.
%The faint end slope is about $\alpha \sim -0.84$, flatter than
%many rich clusters, which might be due to the poor richness of
%A671 and the incompleteness of faint member galaxies.

%%% fig. 9 --------------------------
\begin{figure}
\centering
\includegraphics[height=83mm,width=80mm,angle=0]{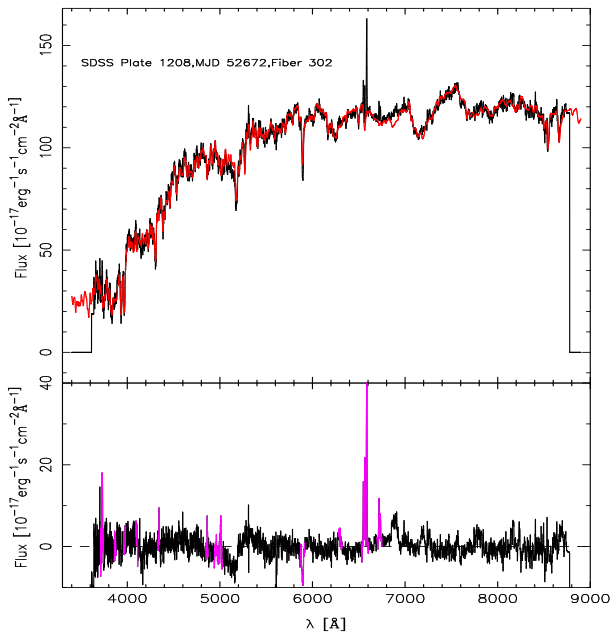}
\caption{Spectral synthesis of the brightest galaxies in A671. Top
panel shows the observed spectrum (black line) and model spectrum
(red line). Bottom panel shows the residual spectrum (black line) and
the mask regions (pink lines). }
\end{figure}

\section{Star Formation History and Element Abundances}

As mentioned in the first section, many former studies support that
galaxy properties strongly correlate with local environment. The
remarkable morphology-segregation of A671 demonstrates that the
morphologies of member galaxies strongly correlate with their local
environment. In this section we would like to investigate the star
formation histories (SFHs) of the confirmed member galaxies by
applying models of stellar population synthesis on their observed
spectra. Two models will be taken in this section. On the one hand,
we fit the SDSS spectra by the STARLIGHT
\footnote{www.starlight.ufsc.br} code \citep{Cid 2005,Mateus 2006,Cid
2007} to derive the physical parameters based on their SFHs. On the
other hand, we compare the absorption line indices (the Lick/IDS
indices) of the member ETGs with the model predictions developed by
\citet{Shiavon 2007} (hereafter S07) to put constraints on the SFHs
and chemical enrichment.

\subsection {Fitting Spectra with STARLIGHT}

We fit the SDSS spectra of member galaxies with the STARLIGHT codes,
which aims at fitting an observed spectra with a liner combination of
theoretical simple stellar populations (SSPs). The model spectrum is
given by
\begin{equation}
M_{\rm \lambda}= M_{\rm \lambda_{0}}(\sum^N_{j=1} x_{\rm j}
b_{\rm j,\lambda}r_{\rm \lambda})\otimes G(v_{\rm *},{\sigma}_{\rm *}),
\end{equation}
where $M_{\rm \lambda}$ is the model spectrum, $M_{\rm \lambda_{0}}$
is the synthesis flux at the normalization wavelength $\lambda_0$,
$x_{j}$ is the so-called population vector, $b_{j,{\lambda}}$ is the
$j$th SSP spectrum at $\lambda$, and
$r_{\lambda}\equiv10^{-0.4(A_{\lambda}-A_{\lambda_{0}})}$ represents
the reddening term.
The $G(v_{*},{\sigma}_{*})$ is the line-of-sight stellar motions that
 modelled by a Gaussian distribution centered at velocity $v_{*}$ and
with a dispersion of ${\sigma}_{*}$.
$N$ is the total number of SSP models. In our work, the SSP base is
made up of $N=45$ SSPs, three metallicities ($Z=0.2 Z_{\odot}$,
$Z_{\odot}$, $2.5Z_{\odot}$) and 15 ages (from 1 Myr to 13 Gyr),
which are taken from evolutionary models in \citet{Bruzual 2003}.
The galactic extinction law of \citet{Cardelli 1989} with
$R_{\rm V}=3.1$ is adopted.

All SDSS observed spectra are shifted to the rest-frame, and then
interpolated into a resolution of 1 ${\textmd{\AA}}$ before fitting.
Wavelength regions of emission lines are masked out. Figure 9 shows
the spectral fitting for the brightest cluster galaxy (BCG) of A671.
As demonstrated by this figure, the combination of SSP spectra can
fit the observed spectrum very well.

The STARLIGHT presents the SSP fraction, intrinsic extinction
$A_{\rm V}$, velocity dispersion $\sigma$, and stellar mass $M_*$.
Following \citet{Cid 2005}, we derive the flux- and mass-weighted
average ages, which are defined as
\begin{equation}
<{\rm log}t_{*}>_{\rm L}=\sum^N_{j=1} x_{\rm j} {\rm log} t_{\rm j};
\,\,\, <{\rm log}t_{*}>_{\rm M}=\sum^N_{j=1} u_{\rm j} {\rm log}
t_{\rm j},
\end{equation}
where $x_{\rm j}$ is the flux-weighted population vector (i.e., the
fraction of flux contributed by certain SSP), and $u_{\rm j}$ is the
mass-weighted population vector.
The average metallicities $<$Z$_{\rm L}$$>$ and $<$Z$_{\rm M}$$>$
can be derived similarly.

\subsubsection{SFH via STARLIGHT Fitting}

\begin{figure*}
\centering
\includegraphics[width=140mm,angle=0]{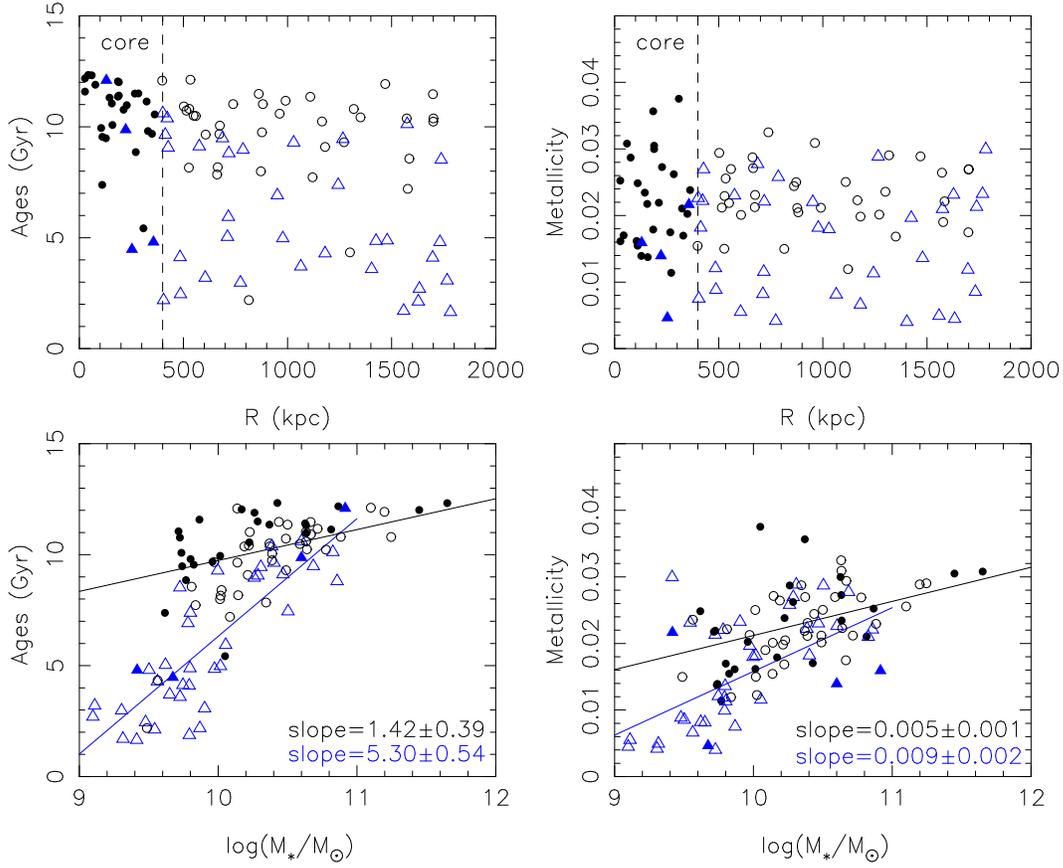}
\caption{Correlation of mass-weighted ages and metallicities with
cluster-centric radius R and stellar mass. Black circles represent
the ETGs, while blue triangles represent the LTGs.
 Galaxies in the core region are denoted by filled symbols. }
\end{figure*}

As illustrated by \citet{Cid 2005}, the individual output vector may
be dramatically deviated from the simulated input value. However, the
average values of stellar age and metallicity should be more
reliable, whatever they are weighted by light or mass. The
flux-weighted age is more sensitive to the young stellar component,
so the mass-weighted age is more underlying and intrinsic. The
situation is the same for average metallicity. In this section, we
will take the average ages and metallicities weighted by stellar
mass. Since stellar population in ETGs are dominated by the old
components, the ETGs have their average ages within a relatively
narrow range.

Figure 10 presents the derived mass-weighted ages and metallicities
as functions of cluster-centric radius R and total stellar mass
($M_{*}$) assembled in cluster galaxies.  The upper two panels show
the mass-weighted ages and metallicities as a function of R. The
galaxies in the core region (R$<$400kpc) of A671 are denoted by
filled symbols. The remarkable morphology-density segregation of A671
is well shown in these two panels. The LTGs (denote by blue symbols)
locate in the outskirts (denoted by open symbols), and have younger
stellar ages. On the other hand, the ETGs with older stellar ages are
located in the core region. For the ETGs of A671, no correlation is
found between metallicities and R.

The lower two panels of Figure 10 present how the mass-weighted ages
and metallicities correlate with stellar mass. Both ages and
metallicities are found to be correlated strongly with stellar mass,
and such correlations are dependent upon morphology. In general, the
more massive galaxies have older ages and richer metallicities. For
the LTGs in A671, the linear correlations of age and metallicity with
stellar mass appear tighter and steeper. Even for the ETGs with
similar stellar mass, the ETGs in core region tend to have older ages
than the outskirt ETGs. However, the ETG metallicities seem to not
vary with cluster-centric radius.

\subsection{Lick Indices}

The age-metallicity degeneracy has haunted stellar population
analysis for decades. Nevertheless, the promising approach to break
it remains the combined use of multiple absorption-line
indices\citep{Kong 2001}. In this section, we will measure the
absorption lines of ETGs in A671 and compare them with state-of-art
SSP models in order to infer their ages, metallicities and
$\alpha$-enhancements.

\subsubsection{Lick Index Measurements}

The bandpasses of the Lick indices are defined in Table 1 of
\citet{Worthey 1994}. We measure the Lick indices with a modified
version of the Lick$\_$EW routine in EZ$\_$Ages package developed
by \citet{Graves 2008b}
\footnote{\textmd{http://www.ucolick.org/${\sim}$graves/EZ$\_$Ages.html}}.
The Lick$\_$EW routine reports the errors of each Lick index
calculated in the way suggested by \citet{Cardiel 1998}. The SDSS
spectral resolution (69 \kms) does not match the originally defined
resolution of Lick indices. The Lick$\_$EW routine smooths the SDSS
spectra to the resolution of Lick/IDS system before measuring the
indices. For the galaxies with high velocity dispersion, the smoothed
absorption features in the SDSS spectra are at resolution even poorer
than the Lick resolution, the Lick$\_$EW routine will apply
$\sigma$-correction for these galaxies. The output includes the
measurements of $\sigma$-corrected indices and their errors.

One of the most challenging issues in Lick index measurements is
emission line contamination. The Balmer absorption features are
contaminated by emission from ionized gas, either from star
formation, AGN activity or interstellar shocks. Balmer line emission
was estimated from equivalent of \ha, which are retrieved from
the  MPA/JHU SDSS DR7 catalog. Some of previous studies used
EW(\hb)$=0.6$ EW(\oiii$\lambda$5007) for correction
\citep{Trager 2000}. We do not use \oiii$\lambda$5007 for correction
because \citet{Nelan 2005} have found that the relation between
\hb\ and \oiii$\lambda$5007 for the ETGs varies in different
mass ranges. They found that the correlation between \hb\ and \ha\
is much tighter. Emission EWs for higher order Balmer lines
are obtained from \ha\ by assuming standard values from Balmer
decrement (in the absence of reddening).
In this way, EW(\hb)$=0.36$ EW(\ha), EW(\hg)$=0.19$ EW(\ha),
and EW(\hd)$=0.13$ EW(\ha).

\subsubsection{SSP Model and Stellar Population Parameters}

Our goal is to use the S07 model to derive the SSP-equivalent
parameters. Firstly we create grid to fit three parameters: ages,
[Fe/H], and [Mg/Fe]. We use the solar-scaled isochrones and the
Salpeter initial mass function suggested by the EZ$\_$Ages documents.
The [O/Fe] is set to be zero, and other $\alpha$ elements are tied
to Mg. The ages
range from 1.2 to 17.7 Gyr, and the [Fe/H] ranges from -1.3 to 0.2.
Other details can be found in S07, \citet{Graves 2008b}.

The three parameters are derived following two steps: i) for each
galaxy, we firstly calculate age and [Fe/H] using \hb\ and
$<$Fe$>$, at all [Mg/Fe]; ii) we compare the pair $<$Fe$>$
($<$Fe$>$=0.5(Fe5270+Fe5335)) and Mgb with models at the median age
obtained in the first step, and calculate a new [Fe/H] and [Mg/Fe].
We then update the ages by interpolating the ages found in the first
step at the [Mg/Fe] derived in the second step, and iterate the
second step with the new age. Usually two iteration steps are needed
before convergence. For those measurements beyond the model grids, we
set the parameters to be the boundaries, i.e, the maximum or minimum
of the models.

Before deriving the stellar population parameters of our sample with
the SSP models, we compare our measurements to the predictions of
each model on the grids of age and metallicity.
In Figure 11, our measurements of \hb\ and [MgFe]$'$ are compared with
the S07 models assuming [$\alpha$/Fe]=0.2, where the index [MgFe]$'$ is
defined as follows:
$[{\rm MgFe}]^{'}$=$\sqrt{{{\rm Mgb}}(0.72{\rm Fe}5270+0.28{\rm Fe}5335)}$.
\textmd{[MgFe}]$'$ is a good indicator of metallicity which is almost
independent upon $\alpha$/Fe ration variations \citep{Thomas 2003}
(hereafter TMB). To convert between [Fe/H] and [Z/H], we adopt the
relation given by TMB: [Z/H]$=$[Fe/H]$+$0.94[$\alpha$/Fe], and we assume
[Mg/Fe]$=$[$\alpha$/Fe]. The median age is about 7 Gyr but with a large
scatter. Most of the core-region galaxies have very small \hb\
values, and occupy the oldest end of age distribution. The [MgFe]$'$
is mainly distributed along [Z/H]$=0$, and does not have any
environmental effects.

%%% fig. 11 -------------------
\begin{figure}
\centering
\includegraphics[width=80mm,angle=0]{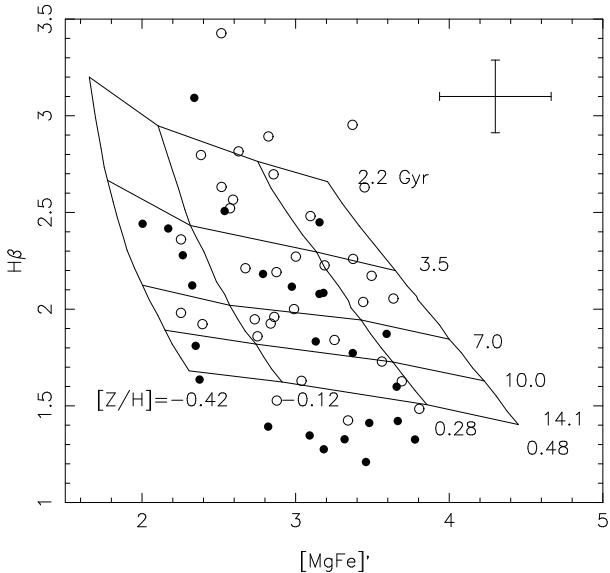}
\caption{
The index plot on the S07 model grid assuming
[$\alpha/$Fe]=0.2, shows the $[$MgFe$]\arcmin$ vs. \hb.
The filled and open circles represent the ETGs in the core
region and those in outskirts, respectively.}
\end{figure}

We present the derived six SSP parameters as a function of velocity
dispersion ($\sigma$) in Figure 12. Galaxies which fall outside model
boundary are not included in the figure. The parameter errors are
given by the fitting procedure for each galaxy. We then compute an
average error for each parameter by weighted the output error of
individual galaxy with its signal-to-noise ratio. As presented in
Figure 11, majority of the galaxies falling out of the model boundary
are located in the core, which makes hard to investigate
environmental effect. Though only 46 ETGs are able to be fitted by
the model, the relations between the SSP parameters and velocity
dispersion ($\sigma$) are still remarkable.

In the top-left panel of Figure 12, the SSP ages show strong
dependence upon velocity dispersion, and the low-$\sigma$ galaxies
span a wider age range, indicating that the low $\sigma$ galaxies
have various possibilities of star formation histories compared with
the high-$\sigma$ ones. Similar results are found in the bright
galaxies in Coma by \citet{Price 2011}. It is noteworthy that the SSP
ages derived by S07 model are not compatible with the average stellar
ages given by the STARLIGHT fitting. Firstly, the SSP ages in
STARLIGHT code range from 1 Myr to 13 Gyr, while the range of the SSP
ages in S07 models is from 1 Gyr to 17.7 Gyr.  The typical age of
ETGs from STARLIGHT fitting (see Figure 10) is older than 7 Gyr. In
Figure 11, the ETG ages derived by the S07 model span a wider range.
Additionally, the STARLIGHT gives the best-fit ages without errors.
The S07 model determines the ages by comparing several
combined-indices with the theoretical model, and the measurement
errors can strongly affect the output ages. For the galaxies whose
indices locate near the model boundary, measurement errors of indices
will bring in greater uncertainties in the parameter fitting.

%%% fig. 12----------------------
\begin{figure*}
\centering
\includegraphics[width=160mm,angle=0]{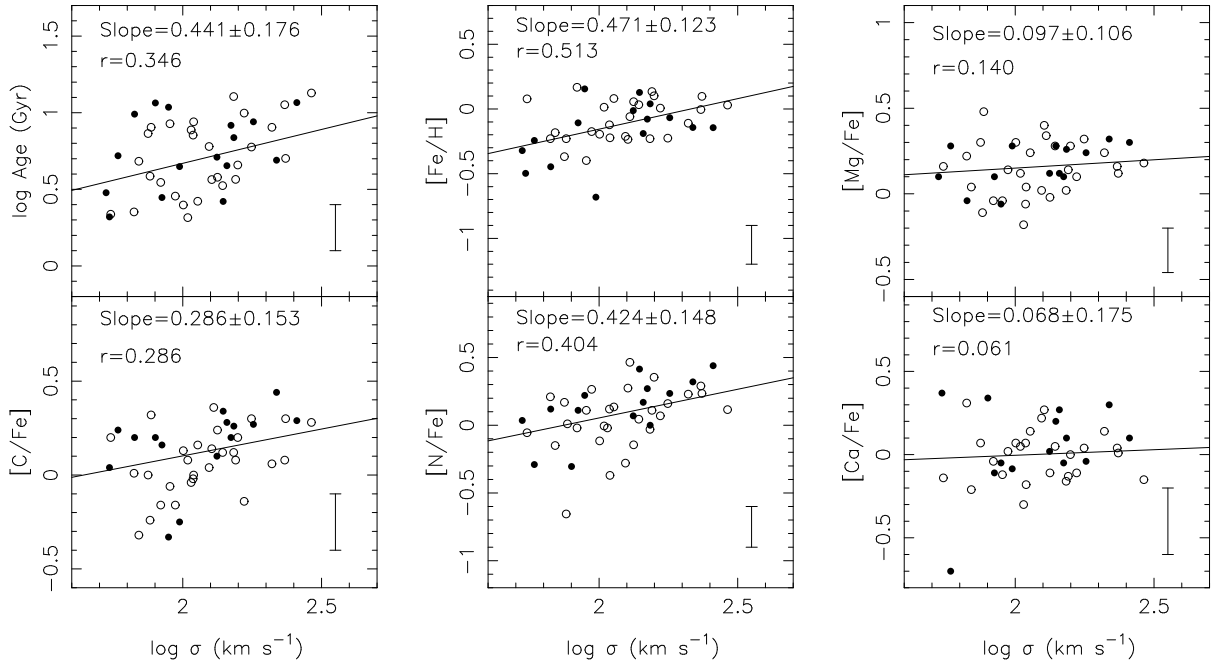}
\caption{ The six SSP-equivalent parameters as the functions
of velocity dispersion $\sigma$ for the 46 galaxies which can
be fitted by the models.
The symbols are defined in Figure 10. The black line represents
the linear fit.}
\end{figure*}

The top-middle and top-right panels display the relation of [Fe/H]
and [Mg/Fe] with $\sigma$. The [Fe/H]$-\sigma$ correlation is very
tight, with a correlation coefficient of $r_{\rm s}=0.513$, whereas
 the [Mg/Fe]$-\sigma$ correlation is rather weak.
The former studies have found [Mg/Fe] strongly correlated with
$\sigma$ \citep{Thomas 2005,Zhu 2010}. We should keep in mind that
the tightness and slope of linear correlation strongly depend on
the sample size. Considering
our small sample size and the intrinsic scatter of this correlation,
a relatively weaker [Mg/Fe]$-\sigma$ relation is still reasonable.

In the three bottom panels of Figure 12, we present [C/Fe]$-\sigma$,
[N/Fe]$-\sigma$, and [Ca/Fe]$-\sigma$ relations. Only a weak
correlation of [N/Fe]$-\sigma$ is found, with a correlation
coefficient of $r_{\rm s}=0.404$.
The results of our fitting are similar
to those in \citet{Graves 2008}, and their results are based on a
large sample of about 6000 red sequence galaxies from the SDSS.

\section{Discussion}

We have investigated the dynamics of A671 based on the
spectroscopically confirmed members, and the result of $\kappa$-test
strongly suggests that A671 has significant substructures. Many
authors have found that a large fraction of galaxy clusters have
substructures \citep{Dressler 1988,Mohr 1993,Yuan 2003,Yang 2004},
indicating that massive clusters may assemble their masses and grow
up by accreting small groups. The contour map of member galaxies in
A671 fits well with the X-ray intensity contour map. \citet{Ramella
2007} suggested that there are two substructures in A671. Location of
one substructure is well associated with the potential substructure
B, and the other is at the south part of cluster, which is not
significant enough to be detected by the $\kappa$-test. Their
substructure-finding algorithm is based on the projected positions of
galaxies, and does not utilize the redshift information.  73\% of the
clusters in their sample were found to have substructures, and this
fraction is higher than most studies. Their magnitude limit of galaxy
samples is $V\simeq$$21^m.2$, much fainter than our limiting
magnitude $h_{\rm BATC}=19^m.5$, thus the projection effect cannot be
ignored, and some substructures they found could be untrue. After 97
newly selected galaxies included, the substructure B is enhanced,
while another substructure appears less prominent. The follow-up
spectroscopy of these faint galaxies is need for revealing the
details of dynamical substructures in A671.

The remarkable morphology-density relation indicates that the cluster
environment indeed have played an important role in evolution of
cluster galaxies. A visual inspection of the C-M diagram shows that a
large fraction of bright member galaxies ($h_{\rm BATC}<18^m.5$) have
evolved to be the ``red sequence'', and faint member galaxies with
$h_{\rm BATC}\sim19^m.0$ are found to have considerable activities of
star
formation. When the galaxies are accreted into a cluster, their star
formation activities are expected to be suppressed by some important
processes, such as tidal stripping and ``harassment'' \citep{Moore
1996}, ram pressure stripping of the gas disk \citep{Abadi 1999}, and
removal of gas reservoir surrounding each galaxy \citep{Balogh 2002}.

We derived the average ages and metallicities for member galaxies by
fitting their spectra. The most remarkable feature of the age
distribution is that the ETGs in the core region have older ages than
those in the outskirts. \citet{Thomas 2005} have found that the ETGs
in dense environment have average ages $\sim 2$ Gyr older than those
in field environment. They derived ages by comparing the Lick indices
with the prediction of TMB model. Our results confirm their
conclusion though the average ages are derived by different methods.
This can be interpreted in two ways. On the one hand, theoretical
work shows that dark matter halos in dense environments were
assembled earlier than average \citep{Gao 2005}. As a result, the
galaxies in the core region of clusters formed earlier than those in
the outer regions, and thus have older stellar ages. On the other
hand, the older stellar age of the ETGs in the core region could be
explained by lack of recent star formation compared with those in the
outer region. Galaxy clusters have dense gas with high temperature,
and the core-region galaxies lost their gas reservoirs by interacting
with the dense intracluster medium (ICM) (usually by ram pressure
stripping). Thus the core-region galaxies should be gas-poor, and
have less possibilities of recent star-formation. Aside from age, the
total metallicities of the ETGs show subtle dependence on
environment, which agrees well with \citet{Zhu 2010}. A larger sample
of cluster ETGs is needed to investigate the environmental effects on
chemical evolution.

The  dependence of age upon stellar mass could be explainable in the
downsizing scenario for galaxy formation \citep{Cowie 1996}. In this
scenario, star formation lasts longer in less massive galaxies than
in more massive galaxies. Thus massive systems will have older SSP
ages on average. Evidences of downsizing effect in the local universe
have been found by many recent studies on the stellar population of
ETGs \citep{Nelan 2005,Graves 2008,Zhu 2010,Price 2011}. The
age-$\sigma$ slope in these studies span a range between 0.35 and
0.93. The different slope values are mainly due to the different
sample property (e.g., galaxy type, sample size, etc.) and methods of
index correction (e.g., emission infill correction, velocity
dispersion correction, etc.). \citet{Price 2011} have tested the
roubustness of their observed age-$\sigma$ slope against these two
factors when they studied the stellar population in Coma. In their
test, a stricter emission-line cut and different methods of velocity
dispersion correction were used. Only slight change on the
age-$\sigma$ slope was found in their tests. They concluded that
their data robustly support the downsizing scenario.

The ETGs also have very strong dependence of $\alpha$-abundance on
$\sigma$. The $\alpha$ elements are mainly from the Type II
supernovae, and the iron-peak elements come mainly from the Type Ia
supernovae. The stronger alpha-enhancement in the more massive
elliptical galaxies may imply that their star formation timescale is
shorter than less massive elliptical galaxies, before the delayed
Type Ia supernovae enrich the star-forming regions with iron-peak
elements. The observed $\alpha$-abundance$-\sigma$ relation fits well
with the prediction from the hierarchical models with feedback
\citep{De Lucia 2006}.

\section{Conclusion}

This paper presents a photometric study of A671 with the
Beijing-Arizona-Taiwan-Connecticut (BATC) multicolor system and the
SDSS data. The main conclusions can be summarized as follow:

(i) About 7000 sources are detected in a BATC field of
$58\arcmin\times58\arcmin$ centered at A671, and their SEDs in 15
intermediate bands are obtained.
The 985 galaxies brighter $h_{\rm BATC}=20^m.0$ are selected by
cross-identifying our BATC source catalog with the released catalog
of SDSS galaxies. There are 205 galaxies with known spectroscopic
redshifts in our viewing field, among which 103 galaxies with
$0.04<\zsp<$0.06 are selected as spectroscopically
confirmed members of A671. The sample of bright member galaxies is
composed of 63 ETGs and 40 LTGs.

(ii) The dynamics of A671 is investigated based on the 103
spectroscopically confirmed members. The result of $\kappa$-test on
different scales strongly suggest that A671 have significant
substructures. Three potential substructures have been suggested with
the method of localized deviation of velocity distribution.

(iii) Photometric redshift technique is applied to the 985 galaxies
for further membership determination. Our photometric redshifts
(\zph) of the bright members are basically consistent with the
spectroscopic redshifts (\zsp). Base on the statistics of
photometric redshifts, the galaxies with $0.028<\zph<0.073$ are
selected as member candidates. After further selection by the
color-magnitude relation, 97 galaxies down to
$h_{\rm BATC}=19^m.5$ are picked up as faint members of A671.

(iv) Based on the enlarged sample of member galaxies, spatial
distribution and velocity structure of A671 are studied. Since the
large \zph\ uncertainty of faint galaxies have smoothed the localized
abnormity in velocity distribution, the $\kappa$-test of the enlarged
sample does not confirm the three substructures mentioned above. The
morphology-segregation becomes very remarkable after the faint
members are taken into account. The luminosity function in the SDSS
r-band shows a flat slope at faint end, $\alpha\sim-1.12$.

(v) Mass-weighted stellar ages and total metallicities of bright
members are derived by fitting their spectra with the spectral
synthesis code, STARLIGHT. The ETGs in the core region have older
ages than those in the outskirts. The more massive ETGs are found to
be older than the less massive ones. No environmental effect is found
for the metallicities of the ETGs. Strong correlations of mean
stellar age and metallicity with stellar mass are confirmed, and such
correlations are found to be dependent upon morphology. The possitive
age-mass correlation supports the downsizing scenario.

(vi) A set of Lick indices of the ETGs is measured in order to
derived their SSP-equivalent stellar parameters (such as age, [Fe/H],
[Mg/Fe], [C/Fe], [N/Fe], and [Ca/Fe]) by utilizing S07 model. The
ETGs at cluster center tend to have smaller \hb\ indices,
indicating that central ETGs are likely to be older. The
total metallicity indicator [MgFe]$'$ does not show any environmental
effects. The relations between the six SSP-parameters and velocity
dispersion ($\sigma$) are also studied. The relations between the
SSP-parameters and $\sigma$ in A671 are in good agreement with
previous studies.

\section*{Acknowledgments}
This work was funded by the National Natural Science Foundation of
China (NSFC) (Grant Nos.~10803007, 10873012, 10873016 and 11173016),
the National Basic Research Program of China (973 Program) (Grant
Nos.~2007CB815403, 2007CB815404), and the Chinese Universities
Scientific Fund (CUSF). This research has made use of the NED, which
is operated by the Jet Propulsion Laboratory, California Institute of
Technology, under contract with the National Aeronautics and Space
Administration.

We acknowledge the use of MPA/JAU Garching DR8 public data. We also
thank Ricardo Schiavon and Genevieve J. Graves for making their
codes and models publicly available.

Funding for the Sloan Digital Sky Survey(SDSS) has been provided by
the Alfred P.Sloan Foundation, the Participating Institutions, the
National Aeronautics and Space Administration, the National Sience
Foundation, the US Department of Energy, the Japanese Monbukagakusho
and the Max-Planck Society. The SDSS web site is http://www.sdss.org.

The SDSS is managed by the Astrophysical Research Consortium for the
participating Institutions. The Participating Institutions are the
University of Chicago, Fermilab, the Institute for Advanced Study,
the Japan Participation Group, Los Alamos National Laboratory, the
Max-Planck Institute for Astronomy(MPIA), New Mexico State
University, University of Pittsburgh, University of Portsmouth,
Princeton Univercity, the United States Naval Observatory and the
University of Washington.


\begin{thebibliography}{99}


\bibitem[\protect\citeauthoryear{Abadi et al.}{1999}]{Abadi 1999}Abadi M. G., Moore B., \& Bower R. G., 1999, MNRAS, 308, 947
\bibitem[\protect\citeauthoryear{Abell}{1958}]{abell58}Abell G. O., 1958, ApJS, 3, 211
\bibitem[\protect\citeauthoryear{Aguerri et al.}{2007}]{Aguerri 2007}Aguerri J. A. L., S$\acute{a}$nchez-Janssen R., \& Mu$\tilde{n}$oz-Tu$\tilde{n}$$\acute{o}$n, 2007, A\&A,
471, 17
\bibitem[\protect\citeauthoryear{Allen}{1976}]{Allen 1976}Allen D. A., 1976, MNRAS, 174, 29
\bibitem[\protect\citeauthoryear{Baldry et al.}{2006}]{Baldry 2006}Baldry I. K., Balogh R. G., Bower K., Glazebrook  K., Nichol R. C., Bamford S. P., \&Budavari T., 2006, MNRAS, 373, 469
\bibitem[\protect\citeauthoryear{Balogh et al.}{2002}]{Balogh
2002}Balogh M. L., et al., 2002, MNRAS, 335, 10
\bibitem[\protect\citeauthoryear{Bautz \& Morgan}{1970}]{bm70}Bautz L. P., \& Morgan, W. W., 1970, ApJL, 162, L149
\bibitem[\protect\citeauthoryear{Beers et al.}{1990}]{Beers 1990}Beers T. C., Flynn K.,
\& Gebhart K., 1990, AJ, 100, 32
\bibitem[\protect\citeauthoryear{Beers et al.}{1991}]{Beers 1991}Beers T. C., Gebhardt K., Forman W., Huchra J. P.,
             Jones C., 1991, AJ, 102, 1581
\bibitem[\protect\citeauthoryear{Beijersbergen et al.}{2002}]{Beijersbergen 2002}Beijersbergen M., Hoekstra H., van Dokkum P. G., \& van de Hulst T., 2002, MNRAS, 329, 385
\bibitem[\protect\citeauthoryear{Bertin \& Arnouts}{1996}]{Bertin 1996}Bertin E. \& Arnouts S., 1996, A\&A, 117, 393
\bibitem[\protect\citeauthoryear{Bolzonella, Miralles \& Pell\'{o}}{2000}]{Bolzonella 2000}Bolzonella M., Miralles J. M., \& Pell\'{o} R., 2000, A\&A, 363,476
\bibitem[\protect\citeauthoryear{Bower et al.}{1992}]{Bower 1992}Bower R. G., Lucey J. R., \& Ellis R. S., 1992, MNRAS, 254, 589
\bibitem[\protect\citeauthoryear{Blanton et al.}{2003}]{Blanton 2003}Blanton M. R. et al., 2003, ApJ, 592, 819
\bibitem[\protect\citeauthoryear{Bruzual \& Charlot}{2003}]{Bruzual 2003}Bruzual G., Charlot S., 2003, MNRAS, 344, 1000
\bibitem[\protect\citeauthoryear{Butcher \& Oemler}{1978}]{Butcher 1976}Butcher H.\& Oemler A. Jr., 1978, ApJ, 226, 559
\bibitem[\protect\citeauthoryear{Cardelli, Clayton \& Mathis}{1989}]{Cardelli 1989}Cardelli J. A., Clayton G. C., Mathis J, S., 1989, ApJ, 345, 245
\bibitem[\protect\citeauthoryear{Cardiel et al.}{1998}]{Cardiel 1998}Cardiel  N., Gorgas J., Cenarro J., \&Gonzalez J. J., 1998, A\&AS, 127, 597
\bibitem[\protect\citeauthoryear{Carlberg et al.}{1997}]{Carlberg 1997}Carlberg R. G., Yee H. K. C., \& Ellingson E., 1997, ApJ, 478, 462
\bibitem[\protect\citeauthoryear{Cid Fernandes et al.}{2005}]{Cid 2005}Cid Fernandes R., Mateus A., Sodr\'{e}, L., Stasi\'{n}ska G., Gomes J. M., 2005, MNRAS, 358, 363
\bibitem[\protect\citeauthoryear{Cid Fernandes et al.}{2007}]{Cid 2007}Cid Fernandes R., Asari N. V., Sodr\'{e}, L., Stasi\'{n}ska G., Mateus A. et al., 2007, MNRAS, 375, L16
\bibitem[\protect\citeauthoryear{Colberg et al.}{2000}]{Colberg 2000}Colberg J. M., White S. D. M. et al., 2000, MNRAS, 319, 209
\bibitem[\protect\citeauthoryear{Colless \& Dunn}{1996}]{Colless 1996}Colless M., \& Dunn A. M., 1996, ApJ, 458, 435
\bibitem[\protect\citeauthoryear{Crone et al.}{1996}]{Crone 1996}Crone M. M., Evrard A. E., \& Richstone D. O., 1996, ApJ,
467,489
\bibitem[\protect\citeauthoryear{Cowie et al.}{1996}]{Cowie 1996}Cowie L. L., Songaila A., Hu E. M., \&cohen J. G., 1996, AJ, 112, 839
\bibitem[\protect\citeauthoryear{De Lucia et al.}{2006}]{De Lucia 2006}De Lucia G., Springel V., White S. D. M., Croton D., \&Kauffmann G. 2006, MNRAS, 366, 499
\bibitem[\protect\citeauthoryear{de Filippis et al.}{2011}]{de Filippis 2011}de Filippis E., Paolillo M., Longo G., La Barbera F., de
Carvalho R. R., Gal R., 2011, MNRAS, 414, 2771
\bibitem[\protect\citeauthoryear{Dressler}{1980}]{Dressler 1980}Dessler A., 1980, ApJ, 236, 351
\bibitem[\protect\citeauthoryear{Dressler et al.}{1997}]{Dressler 1997}Dessler A., Oemler A. Jr. et al., 1997, ApJ, 490, 577
\bibitem[\protect\citeauthoryear{Dressler}{1988}]{Dressler 1988}Dressler A. \&
Shectman S. A., 1988, AJ, 95, 985
%\bibitem[\protect\citeauthoryear{Driver et al.}{1994}]{Driver 1994}Driver S. P., Philipps S., Davies J. I., Morgan I.,\& disney M. J., 1994,
%MNRAS, 268, 393
\bibitem[\protect\citeauthoryear{Ebeling et al.}{1998}]{Ebeling 1998}Ebeling H., Edge A. C., B$\ddot{o}$hringer H., Allen S. W.,
Crawford C. S., Fabian A. C., Voges W., Huchra J. P., 1998,
MNRAS,301, 881
\bibitem[\protect\citeauthoryear{Fan et al.}{1996}]{Fan 1996}Fan X., Burstein D., Chen J. S. et al., 1996, AJ, 112, 628
\bibitem[\protect\citeauthoryear{Fasano et al.}{2000}]{Fasano 2000}Fasano G., Poggianti B. M., Couch W. J., Bettoni D. et al., 2000, ApJ, 542, 673
\bibitem[\protect\citeauthoryear{Fern\'{a}ndez-Soto et al.}{1999}]{Fernadez 1999}Fern\'{a}ndez-Soto A., Lanzetta K. M., Yahil A., 1999, ApJ, 513, 34
\bibitem[\protect\citeauthoryear{G\'{o}mez et al.}{2003}]{Gomez 2003}G\'{o}mez P. L., Nichol R. C., Miller C. J. et al., 2003, ApJ, 584, 210
\bibitem[\protect\citeauthoryear{Gao et al.}{2005}]{Gao 2005}Gao L., Springel V. \& White S. D. M., 2005, MNRAS, 363, L66
\bibitem[\protect\citeauthoryear{Geller \& Peebles}{1973}]{Geller 1973}Geller M. J. \& Peebles P. J. E., 1973, ApJ, 184, 329
\bibitem[\protect\citeauthoryear{Gott}{1972}]{Gott 1972}Gott J. R. I., 1972, ApJ, 173, 277
\bibitem[\protect\citeauthoryear{Goto et al.}{2003}]{Goto 2003}Goto T., Tamauchi C., Fujita Y., Okamura S. et al., 2003, MNRAS, 346, 601
\bibitem[\protect\citeauthoryear{Graves et al.}{2008}]{Graves 2008}Graves G. J., Faber S. M., Schiavon R. P., Yan R., 2007, ApJ, 671,
243
\bibitem[\protect\citeauthoryear{Graves \& Schiavon}{2008}]{Graves 2008b}Graves G. J., \& Schiavon R. P., 2008, ApJS, 177, 446
%\bibitem[\protect\citeauthoryear{Gonz\'{a}lez}{1993}]{Gonzalez 1993}Gonz\'{a}lez J. J., 1993, PhD thesis, Thesis(PH.D)-UNIVERSITY OF CALIFORNIA, SANTA CROUZ, 1993. Source:Dissertation Abstracts International, Volume:54-05, Section: B, page:2551
\bibitem[\protect\citeauthoryear{Gunn \& Stryker}{1983}]{Gunn 1983}Gunn J. E., \& Stryker L. L., 1983, ApJS, 52, 121
\bibitem[\protect\citeauthoryear{Hansen et al.}{2005}]{Hansen 2005}Hansen S. M., Mckay T. A., Wechsler R. H., Annis J., Sheldon E. S., \&Kimball A., 2005, ApJ, 633, 122
\bibitem[\protect\citeauthoryear{Holden et al.}{2007}]{Holden 2007}Holden B. P., Illingworth G. D., Franx M., Blakeslee J. P. et al., 2007, ApJ, 670, 190
\bibitem[\protect\citeauthoryear{Ilbert et al.}{2009}]{Ilbert 2009}Ilbert O., Capak P.,
 Salvato M. et al., 2009, ApJ, 690, 1236
\bibitem[\protect\citeauthoryear{Jones \& Forman}{1999}]{Jones 1999}Jones C. \& Forman W., 1999, ApJ, 511,65
\bibitem[\protect\citeauthoryear{Kauffmann et al.}{2004}]{Kauffmann 2004}Kauffmann G., White S. D. M., Heckman T. M. et al., 2004, MNRAS, 353, 713
\bibitem[\protect\citeauthoryear{Kong et al.}{2000}]{Kong 2000}Kong X., et al., 2000, AJ, 119, 2745
\bibitem[\protect\citeauthoryear{Kong \& Cheng}{2001}]{Kong 2001}Kong X., \& Cheng F.Z., 2001, MNRAS, 323, 1035
\bibitem[\protect\citeauthoryear{Kong et al.}{2009}]{Kong 2009}Kong X., Fang G., Arimoto N., \& Wang M.\ 2009, ApJ, 702, 1458
\bibitem[\protect\citeauthoryear{Liu et al.}{2011}]{Liu 2011}Liu
S. F., Yuan Q. R., Yang Y. B., Ma J., Jiang Z. J., Wu J. H., Wu Z.
Y., Chen J. S. \& Zhou X., 2011, AJ, 141, 99
\bibitem[\protect\citeauthoryear{Mateus et al.}{2006}]{Mateus 2006}Mateus A., Sodr\'{e} L., Cid Fernandes R. et al., 2006, MNRAS, 370, 721
\bibitem[\protect\citeauthoryear{Mohr et al.}{1993}]{Mohr 1993}Mohr J. J.,
Fabricant D. G. \& Geller M. J., 1993, ApJ, 413, 492
\bibitem[\protect\citeauthoryear{Moore et al.}{1996}]{Moore 1996}Moore B.,
Katz N., Lake G., Dressler A., \& Oemler A., 1996, Nature, 379, 613
\bibitem[\protect\citeauthoryear{Muratov \& Gnedin}{2010}]{Muratov 2010}Muratov A. L. \& Gnedin O. Y., 2010, ApJ, 718, 1266
\bibitem[\protect\citeauthoryear{Nelan et al.}{2005}]{Nelan 2005}Nelan J. E., Smith R. J., Hudson M. J., Wegner G. A., Lucey J. R. et al., 2005, ApJ, 632, 137
\bibitem[\protect\citeauthoryear{Oegerle \& Hill}{1994}]{Oegerle
1994}Oegerle W. R., \& Hill J. M., 1994, AJ, 107, 857
%\bibitem[\protect\citeauthoryear{Poggianti}{2004}]{Poggianti 2004} Poggianti B., 2004, Proceedings of Baryons in Dark Matter Halos, Novigrad, Croatia, 5-9 Oct 2004, eds. R. Dettmar, U. Klein, \& P. Salucci (SISSA, Proceedings of Science), 104.1
\bibitem[\protect\citeauthoryear{Paolillo et al.}{2001}]{Paolillo
2001}Paollilo M., Andreon S., Longo G., Puddu E., Gal R. R.,
Scaramella R., Djorgovski S. G., de Carvalho R., 2001, A\&A, 367, 59
\bibitem[\protect\citeauthoryear{Postman \& Geller}{1984}]{Postman 1984}Postman M., \& Geller M. J., 1984, ApJ, 281, 95
\bibitem[\protect\citeauthoryear{Postman et al.}{2005}]{Postman 2005}Postman M. et al., 2005, ApJ, 623, 721P
\bibitem[\protect\citeauthoryear{Price et al.}{2011}]{Price 2011}Price J., Phillipps S., Huxor A., Smith R. J. \& Lucey J. R., 2011, MNRAS, 411, 2558
\bibitem[\protect\citeauthoryear{Ramella et al.}{2007}]{Ramella 2007}Ramella M., Biviano A., Pisani A., Varela J., Bettoni D. et al., 2007, A\&A, 470, 39
\bibitem[\protect\citeauthoryear{Rhee et al.}{1991}]{Rhee 1991}Rhee G. F. R. N.,
van Haarlem M. P., Katgert P., 1991, A\&A, 233,325
\bibitem[\protect\citeauthoryear{Shiavon}{2007}]{Shiavon 2007}Schiavon R. P., 2007, ApJS, 171, 146
\bibitem[\protect\citeauthoryear{Schechter}{1976}]{Schechter 1976}Schechter P., 1976, ApJ, 203, 297
\bibitem[\protect\citeauthoryear{Schuecker et al.}{2001}]{Schuecker 2001}Schuecker P., B\"{o}hringer H., Reiprich T. H, \& Feretti L., 2001 A\&A, 378, 408
%\bibitem[\protect\citeauthoryear{Smith et al.}{1997}]{Smith 1997}Smith R. M., Driver S. P. \& Phillipps S., 1997, MNRAS, 287, 415
\bibitem[\protect\citeauthoryear{Smith et al.}{2005}]{Smith 2005}Smith G. P., Treu T., Ellis R. S., Moran S. M., Dressler A.,
2005, ApJ, 620, 78S
\bibitem[\protect\citeauthoryear{Thomas et al.}{1998}]{Thomas
1998}Thomas P. A. et al., 1998, MNRAS, 296, 1061
\bibitem[\protect\citeauthoryear{Thomas et al.}{2003}]{Thomas 2003}Thomas D., Maraston C. \&Bender R., 2003, MNRAS, 339, 897
\bibitem[\protect\citeauthoryear{Thomas et al.}{2005}]{Thomas 2005}Thomas D., Maraston C., Bender R., Mendes de Oliveira C., 2005, ApJ, 621, 673
\bibitem[\protect\citeauthoryear{Trager et al.}{2000}]{Trager 2000}Trager S. C., Faber S. M., Worthey G. \& Gonz\'{a}lez J. J., 2000, AJ, 119, 1645
\bibitem[\protect\citeauthoryear{West et al.}{1991}]{West 1991}West M. J., Villumsen J. V., Dekel A., 1991, ApJ, 369, 287
\bibitem[\protect\citeauthoryear{West et al.}{1995}]{West 1995}West M. J., Jones C., Forman W., 1995, ApJ, 451, L5
\bibitem[\protect\citeauthoryear{Whitmore \& Gilmore}{1991}]{Whitmore 1991}Whitmore B. C. \& Gilmore D. M., 1991, ApJ, 367, 64W
\bibitem[\protect\citeauthoryear{Worthey et al.}{1994}]{Worthey 1994}Worthey G., Faber S. M., Gonzalez J. J. \& Burstein D., 1994, ApJS, 94, 687
\bibitem[\protect\citeauthoryear{Xia et al.}{2002}]{Xia 2002} Xia L. F., Zhou X., Ma J.,et al., 2002, PASP, 114, 1349
\bibitem[\protect\citeauthoryear{Yang et al.}{2004}]{Yang 2004}Yang Y. B., Zhou X., Yuan Q. R., Jiang Z. J.,
 Ma J., Wu H., Chen,J. S., 2004, ApJ, 600, 141
\bibitem[\protect\citeauthoryear{Yuan et al.}{2003}]{Yuan 2003}Yuan Q. R., Zhou X., Jiang Z, J., 2003, ApJS,
149, 53
\bibitem[\protect\citeauthoryear{Yuan et al.}{2001}]{Yuan 2001}Yuan Q. R., Zhou X., Chen J. S., Ma J., Wu H.,
Xue S. J., Zhu J., 2001, AJ, 122, 1718
%\bibitem[\protect\citeauthoryear{Zeldovich et al.}{1982}]{Zeldovich 1982}Zeldovich I. B., Einasto J., Shandarin S. F., 1982, Nature, 300, 407
%\bibitem[\protect\citeauthoryear{Zhang et al.}{2010}]{Zhang 2010}Zhang L., Yuan Q. R., Zhou X., et al., 2010, RAA, 10,
%1
\bibitem[\protect\citeauthoryear{Zhang et al.}{2011}]{Zhang 2011}Zhang L., Yuan Q. R., Yang Q., et al., 2011, PASJ, 63,
585
\bibitem[\protect\citeauthoryear{Zhou et al.}{1999}]{Zhou 1999}Zhou X., Chen J. S., Xu W. et al., 1999, PASP, 111, 909
\bibitem[\protect\citeauthoryear{Zhou et al.}{2001}]{Zhou 2001}Zhou X., Jiang Z. J.,
Xue S. J., Wu H., Ma J., Chen J. S., 2001, ChJAA, 1, 372
%\bibitem[\protect\citeauthoryear{Zhou et al.}{2003}]{Zhou 2003}Zhou X., Jiang Z. J.,
%Ma J. et al., 2003, A\&A, 397, 361
\bibitem[\protect\citeauthoryear{Zhu et al.}{2010}]{Zhu 2010}Zhu G. T., Blanton M. R. \& Moustakas J., 2010, ApJ, 722, 491



\end{thebibliography}
\end{document}